Balance of $O_2$ and $H_2$ content under laser-induced breakdown of aqueous colloidal solutions

E.V.Barmina[1], A.V. Simakin[1], and G.A. Shafeev[1,2*]

[1] Wave Research Center of A.M. Prokhorov General Physics Institute of the Russian Academy of Sciences, 38, Vavilov street, 119991 Moscow Russian Federation

[2] National Research Nuclear University MEPhI (Moscow Engineering Physics Institute), 31, Kashirskoye highway, 115409, Moscow, Russian Federation

* Corresponding author, e-mail: shafeev@kapella.gpi.ru

**Abstract**

Formation of molecular $H_2$ and $O_2$ is experimentally studied under laser exposure of water colloidal solution to radiation of a Nd:YAG laser at pulse duration of 10 ns and laser fluence in the liquid of order of 100 J/cm$^2$. It is found the partial pressure of both $H_2$ and $O_2$ first increases with laser exposure time and saturates at exposures of order of 1 hour. The balance between $O_2$ and $H_2$ content depends on the laser energy fluence in the solution and is shifted towards $H_2$ at high fluences. Possible mechanisms of formation of the dissociation products are discussed, from direct dissociation of $H_2O$ molecules by electrons of plasma breakdown to emission of laser-induced plasma in liquid.

**Introduction**

Laser ablation of solids in liquids is a physical method of generation of large variety of nanoparticles. Typically, a solid target is placed into liquid, which is transparent for laser radiation. If the laser fluence on the target is enough for its melting, then this melt is dispersed into surrounding liquid as nanoparticles (NPs). Virtually any liquid is suitable for generation of NPs in this way, water being the most common of them. Organic solvents, such as alcohols, are also frequently used for preparation of colloidal solutions of different NPs in them. Laser ablation of solids in liquids is accompanied by formation of plasma plume above the target. Plasma is also observed under laser exposure of colloidal solutions of NPs. If laser intensity is high enough for NPs to reach temperatures of about $10^4$-$10^5$ K, some part of their atoms may be ionized [1]. Most of the liquids used for laser generation of NPs contain hydrogen or its isotopes. In conditions of laser-induced breakdown both the liquid and the material of NPs are affected resulting in chemical changes of their composition. In case of organic solvents, e.g., ethanol, this leads to deep pyrolysis of the liquid down to formation of elementary glassy carbon [2]. One should also expect the formation of gaseous products of liquid decomposition. Indeed, as it was recently shown, laser exposure of colloidal solution of nanoparticles in water is accompanied by emission of molecular $H_2$ [3]. There should be other gaseous products of $H_2O$ molecules decomposition, for example, molecular $O_2$. In this work we present new results concerning the formation of both molecular $O_2$ and $H_2$ simultaneously and independently measured under laser exposure of $H_2O$ containing small nanoparticles. Different pathways of formation of these products are discussed.

**Experimental technique**

Experimental setup is shown in Fig. 1. We used technical water that contained about $10^{11}$ $cm^{-3}$ particles with average size of 2.5 nm (as determined with measuring disk centrifuge) and made mostly of iron oxides and hydroxides. Qualitatively the results remain the same using the purest $H_2O$ available (conductivity of 0.6 μSm) though the presence of nanoparticles leads to higher rate of gases production.

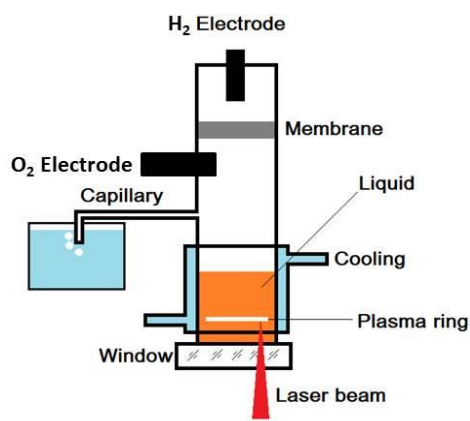

Fig.1. Experimental setup with two amperometric sensors for independent measurements of both $H_2$ and $O_2$ content in the atmosphere above the liquid.

The role of nanoparticles is to ignite the primary breakdown of the colloidal solution which then develops into microscopic breakdown channel of the liquid. Further irradiation of NPs colloidal solutions in absence of the target and diluted in necessary proportion was carried out using the radiation of the Nd:YAG laser at wavelength of 1064 nm and pulse duration of about 10 ns (FWHM). Laser radiation was focused inside the liquid by an F-Theta objective with focal distance of 90 mm (Fig. 1). Laser beam was scanned across the window along circular trajectory about 8 mm in diameter at the velocity of 1000 – 3000 mm/s by means of galvo mirror system. Laser exposure of 4 ml portions of colloids was carried out at 2 mJ energy per pulse and repetition rate of laser pulses of 10 kHz. Estimated diameter of the laser beam waist was 30 μm, which corresponds to laser fluence in the liquid up to 125 J/cm$^2$. Bright cylinder made of plasma appeared 2-3 mm above the window inner surface and looked continuous for eye.

Amperometric molecular hydrogen sensor was used to monitor the concentration of $H_2$ in the space above the liquid surface. Excessive pressure in the system was released to ambient air through a glass capillary dipped in 2 mm thick water layer. In this case the total pressure in the cell was equal to atmospheric one. Inner electrolyte of the sensor is separated from the cell atmosphere by a membrane pervious only to $H_2$. The sensor indicates either the concentration of $H_2$ in μg/l (mg/l) or its partial pressure in Torrs. Calibration of the sensor was performed in air (no $H_2$) and in 1 atmosphere of $H_2$. The atmosphere pressure was checked for each day of measurements since is changes with time. The precision of $H_2$ concentration measurements is 5%. Total volume of the atmosphere above the water level can be estimated as 10 ml. The characteristic time of sensor response in this geometry is about 5 min.

Molecular oxygen sensor, also amperometric one, was attached to the atmosphere above the liquid. Initial content of $O_2$ was set to 20% at atmosphere pressure, and then the calibration was performed automatically according to sensor software. Both sensors allowed independently monitor the content of $O_2$ and $H_2$ in the atmosphere above the irradiated liquid during and after laser exposure. Moreover, both sensors were used for measurements of $H_2$ and $O_2$ content in the liquid right after laser exposure.

**Results**

Hydrogen and oxygen yields dependence of exposure time and laser fluence in aqueous colloidal solution are presented in Fig. 2, a,b. The interesting phenomenon consists in the dependence of $O_2$ concentration on laser exposure time. $O_2$ concentration decreases at the beginning of laser exposure and only then starts to grow. On the contrary, $H_2$ concentration only increases with time of exposure. This difference is probably due to the fact that the initial $O_2$ concentration is around 20%. After the onset of laser exposure emitted $H_2$ pushes initial $O_2$ out of measuring volume. Then the $O_2$ concentration starts to increase due to emission of $O_2$ that originates from $H_2O$ dissociation.

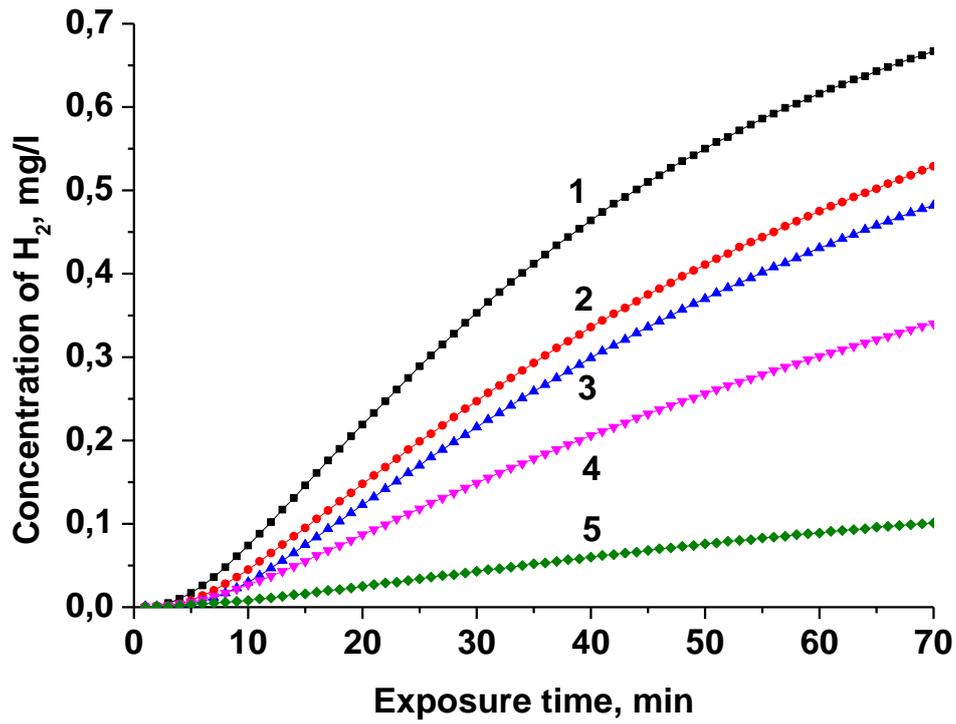

Fig. 2,a. Concentration of $H_2$ in atmosphere above the liquid as the function of time of laser exposure and laser fluence in the liquid, 1 -125, 2 – 124, 3 – 123, 4 – 118, and 5 – 110 J/cm$^2$.

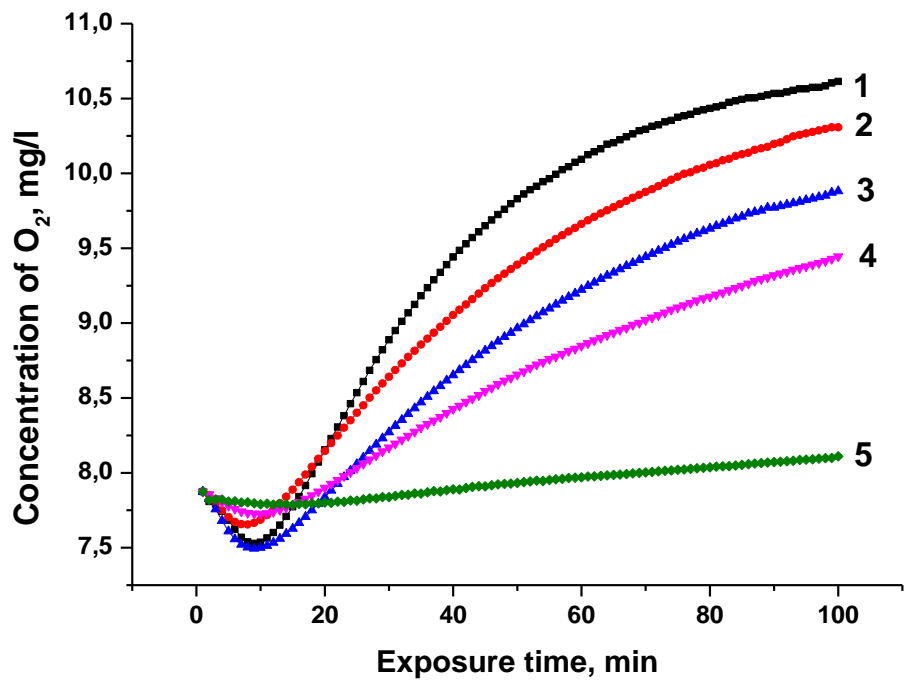

Fig. 2,b. Concentration of $O_2$ in atmosphere above the liquid as the function of time of laser exposure and laser fluence in the liquid, 1 -125, 2 – 124, 3 – 123, 4 – 118, and 5 – 110 J/cm$^2$.

In other words, the liquid upon laser exposure contains explosive mixture of $H_2$ and $O_2$. From time to time the microscopic bubbles containing both gases are formed. If such a bubble crosses with laser beam upon its scanning then well discernable micro-explosion is heard. In some cases, however, these micro-explosions are strong enough to provoke irreversible damage of the glass cell.

It should be noted that the sum of partial pressures of $H_2$ and $O_2$ (640 Torr) is lower than the atmospheric pressure of 750 Torr. This suggests the presence of another gaseous product of $H_2O$ decomposition that could not be detected with amperometric sensors used.

The lower is the laser fluence, the more time is needed to reach the saturation level of $H_2$. In general, however, it was found that the temporal dependence of gas pressure on exposure time follows sigmoidal dependence (see Figs. 2 a, b). Then the $H_2$ content at relatively low laser fluence (<140 $J/cm^2$) was approximated by such a curve, and the pressure was taken to that reached at infinitely long laser exposure time.

One can see that the relative content of $H_2$ shifts towards hydrogen with the increase of laser fluence. Indeed, there is no $H_2$ before the onset of laser exposure, while $O_2$ content corresponds to that one in ambient atmosphere (20%).

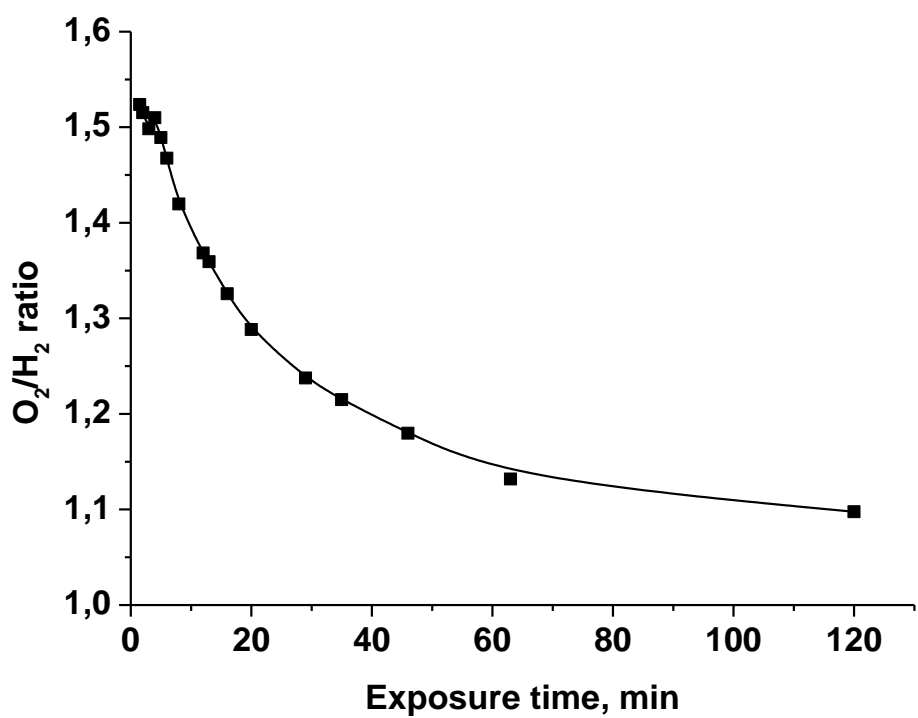

Fig. 3. Dependence of $O_2/H_2$ ratio in number of molecules on the exposure time. Laser fluence of 120 $J/cm^2$.

The balance between $H_2$ and $O_2$ also depends on the laser fluence inside the liquid at otherwise equal experimental conditions. Fig. 4 presents the dependence of this ratio on the laser fluence. At the threshold of plasma channel formation $H_2$ content and $H_2/O_2$ ratio are close to 0. At higher fluences $H_2/O_2$ ratio increases and reaches the plateau value of 1.4 at fluence above 130 $J/cm^2$. This value apparently corresponds to the equal rate of $H_2$ and $O_2$ emission.

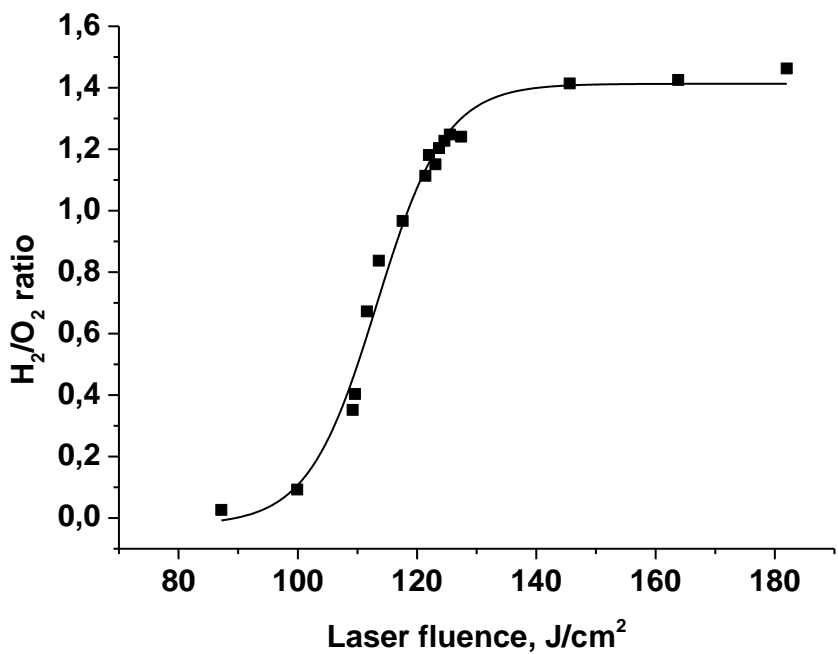

Fig. 4. Ratio $H_2/O_2$ in number of molecules as the function of laser fluence in the liquid.

It should be stressed once again that the stationary levels of gases content at low laser fluence were obtained from sigmoidal approximation of the initial dependences.

**Dependence of content of both $H_2$ and $O_2$ in aqueous medium and in atmosphere above it**

Upon the onset of $H_2$ emission the liquid should first be saturated with it and only then will penetrate into measuring volume. This does not concern, of course, relatively large gas bubbles that ascend much faster. Both amperometric sensors for $H_2$ and $O_2$ allow measuring the content of corresponding gases either in the liquid or in the atmosphere about it. For measurements of gas content inside the liquid the corresponding sensor was dipped into the liquid right after laser exposure.

Concentration of $H_2$ and $O_2$ in the colloidal solution and above it as a function of exposure time is shown in Figs. 5 a, b. One can see that the concentration of $H_2$ in liquid monotonously increases with exposure time. Same concerns the concentration of $O_2$ but the latter starts from the initial relatively high value, so the initial depression of $O_2$ concentration in ambient atmosphere above the liquid is clearly visible.

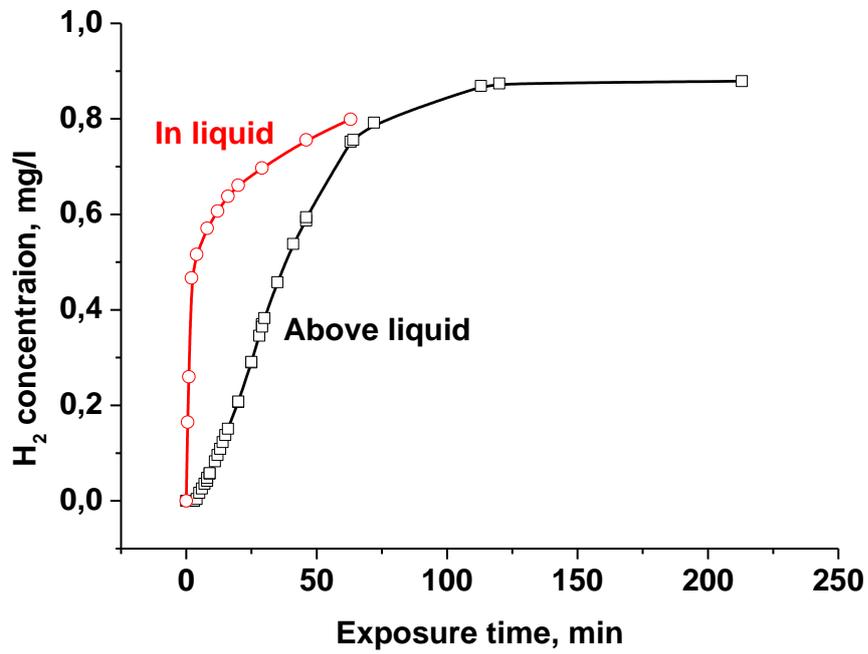

Fig. 5, a

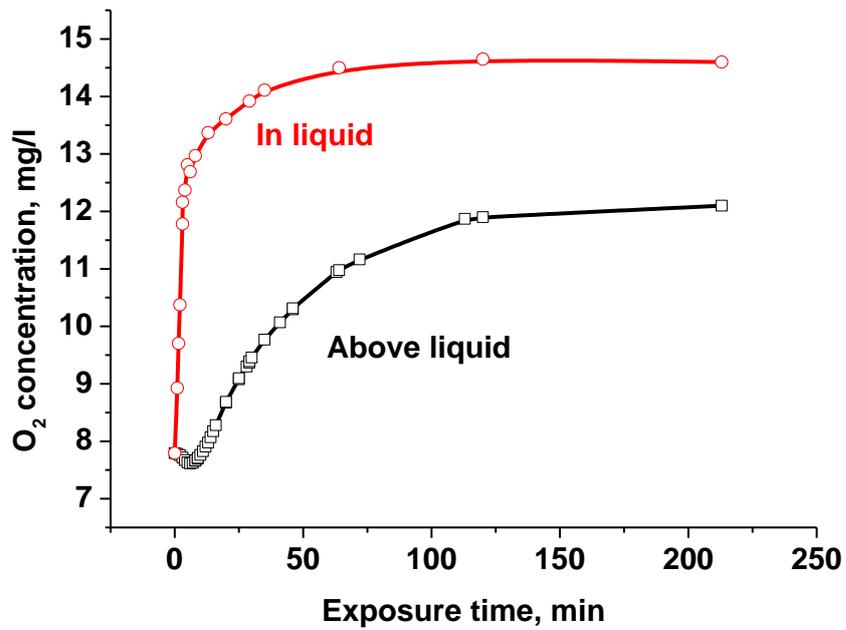

Fig. 5, b

Fig. 5. Dependence of concentration of $H_2$ (a) and $O_2$ (b) in liquid (circles) and above the liquid (squares) on laser exposure time.

At first sight one may conclude that most of $O_2$ remains in the liquid due to lower diffusion coefficient compared to $H_2$. However, long term measurements of gaseous content during 5 hours after laser exposure show that finally $O_2$ diffuses into the atmosphere above the liquid, so that its content above the liquid becomes equal to that in the liquid (Fig. 6).

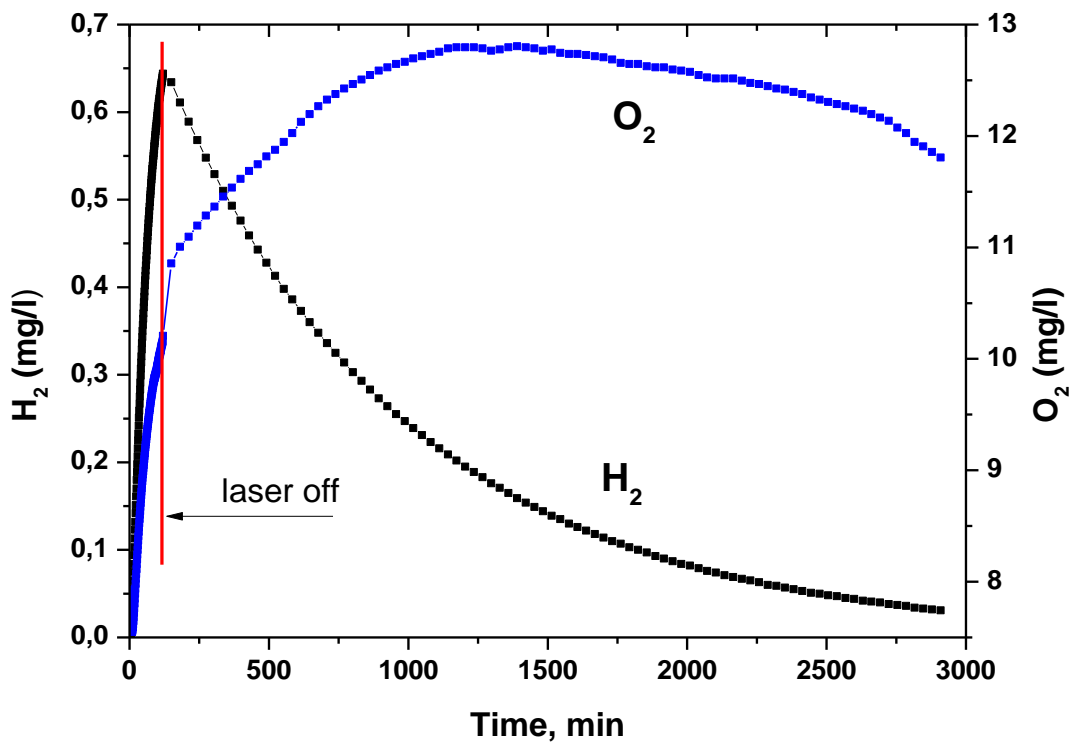

Fig. 6. Long term evolution of $H_2$ and $O_2$ content. The vertical line indicates the moment of switching laser off. Laser fluence of 105 J/cm$^2$.

On the contrary, $H_2$ escapes from measuring volume due to diffusion to surrounding atmosphere trough leaks in the cell.

**Discussion**

The dependence of concentration of both $H_2$ and $O_2$ (Fig. 2) shows threshold-like character on the laser fluence inside the liquid. Indeed, the rate of gases generation decreases with the decrease of fluence. Qualitatively, similar behavior is also observed for the brightness of breakdown plasma induced in the liquid. Therefore, there is a positive feedback in the system "laser radiation – plasma". The energy of laser photons of approximately 1 eV is not sufficient to induce the breakdown of $H_2O$ molecules. The probability of multi-photon absorption of laser quanta is also low due to relatively long laser pulses of order of 10 ns. On the other hand each nanoparticle in the laser beam waist is a source of so called "nanoplasma" [4]. At sufficiently high laser fluence these nanoplasmas unite in microscopic channel directed along laser beam waist.

The most probable reaction channels that lead to emission of both $O_2$ and $H_2$ under laser exposure of colloidal water solutions are dissociation of $H_2O$ molecules by direct electron impact from the laser plasma breakdown. As soon as microscopic plasma channel is formed, there is no liquid $H_2O$ in it, and it is filled with $H_2O$ vapors.

Most probably the dissociation of $H_2O$ molecules occurs just in this channel in $H_2O$ vapors, because the density of vapors is much lower than that of the liquid. Therefore, electrons of breakdown plasma may acquire sufficient energy for $H_2O$ dissociation from laser beam due to higher mean free path. Some products that appear under collision of electron beam with $H_2O$ molecules relevant to formation of oxygen with hydrogen are as follows [5]:

$H_2^+ + O + 2e$ (Eap = 20.70 eV);

$O^+ + H_2 + 2e$ (Eap = 19.00 eV);

$O^+ + 2H + 2e$ (Eap = 26.80 eV)

Here the symbol 'ap' stands for appearance of corresponding process. The shift of $H_2/O_2$ balance with variation of laser fluence (Fig. 3) should be interpreted then as variation of electronic temperature in the plasma channel with laser fluence. This explains the deviation of $H_2/O_2$ balance from stoichiometric value (Figs. 3 and 4).

**Conclusion**

Thus, laser-induced breakdown of aqueous colloidal solutions at energy density of order of 100 J/cm$^2$ is accompanied by emission of both $H_2$ and $O_2$. The balance between these gases depends on the energy density in the solution. Emission of both gases is ascribed to direct dissociation of $H_2O$ molecules by electrons of breakdown plasma. Another possibility consists in dissociation of $H_2O$ molecules by high-energy photons that are generated inside the plasma channel. This radiation is Bremsstrahlung and has rather wide spectrum. High-energy tails of photon spectra may contribute to dissociation of $H_2O$ molecules giving rise to formation of $H_2$ and $O_2$ in present experimental conditions. The contribution of photons of laser-induced plasma to emission of $H_2$ and $O_2$ requires further studies. Finally, the high-energy photons may come from collapsing gas bubbles. The relative role of these factors on $H_2O$ dissociation requires further studies. Formation of other products, such as $H_2O_2$ and $O_3$ may also take place but this cannot be detected with gas sensors used in present work. The presented results point out to the fact that both $H_2$ and $O_2$ are produced almost in any experiment on laser ablation in liquids, such as water and alcohols. The emission of these gases may alter the chemical composition of generated nanoparticles with respect to the material of the initial target.

**Acknowledgements**

The authors gratefully acknowledge the support of the Russian Foundation for Basic Researches, Grants 15-02-04510_a, 15-32-20926_mol_a_ved, 16-02-01054_a, and RF President's Grant MK-4194.2015.2.